\documentclass[conference]{IEEEtran}
\IEEEoverridecommandlockouts
\usepackage{cite}
\usepackage{amsmath,amssymb,amsfonts}
\usepackage{graphicx}
\usepackage{textcomp}
\usepackage{xcolor}
\def\BibTeX{{\rm B\kern-.05em{\sc i\kern-.025em b}\kern-.08em
    T\kern-.1667em\lower.7ex\hbox{E}\kern-.125emX}}
    
\usepackage{graphicx}
\usepackage{amsmath}
\usepackage{amssymb}
\usepackage[caption=false]{subfig}
\usepackage{float}
\usepackage{bibentry}
\usepackage{balance}
\usepackage{algorithmicx}
\usepackage{algpseudocode}
\usepackage{xcolor}
\IEEEoverridecommandlockouts
\graphicspath{{img/}}
\usepackage[top=0.75in, left=0.625in, bottom=1.0in, right=0.625in]{geometry}

\begin{document}
\title{ 
 Lifetime Maximization for UAV-assisted  Data Gathering  Networks in the Presence of Jamming}

\author{Ali Rahmati$^*$, Seyyedali Hosseinalipour$^*$,  \.{I}smail G\"{u}ven\c{c}$^*$, Huaiyu Dai$^*$, and Arupjyoti Bhuyan$^\dagger$\\
$^*$Department of Electrical and Computer Engineering, North Carolina State University, Raleigh, NC\\
$^\dagger$Idaho National Laboratory, Idaho Falls, ID\\
Email:{\tt  \{arahmat,shossei3,iguvenc,hdai\}@ncsu.edu, arupjyoti.bhuyan@inl.gov}\thanks{This work is supported in part through the INL Laboratory Directed Research \& Development (LDRD) Program under DOE Idaho Operations Office Contract DE-AC07-05ID14517.}
}

\maketitle

\begin{abstract}
Deployment of unmanned aerial vehicles (UAVs) is recently getting significant attention due to a variety of practical use cases, such as surveillance, data gathering, and commodity delivery. Since UAVs are powered by batteries, energy  efficient communication is of paramount importance.  In this paper, we investigate the problem of lifetime maximization of a UAV-assisted  network  in the presence of multiple sources of interference, where the UAVs are deployed to collect data from a set of wireless sensors. We demonstrate that the placement of the UAVs play a key role in prolonging the lifetime of the network  since the required transmission powers of the UAVs are closely related to their locations in space. In the proposed scenario,  the UAVs transmit  the gathered data to a primary UAV called \textit{leader}, which is in charge of forwarding the data to the  base station (BS) via a backhaul UAV network. We deploy tools from spectral graph theory  to tackle the problem due to its high non-convexity. Simulation results demonstrate that our proposed method can significantly improve the lifetime of the UAV network.
\end{abstract}


\section{Introduction}
\noindent The utilization of unmanned aerial vehicles (UAVs) is a new concept with a vast majority of applications, such as video monitoring and surveillance, package delivery, and help and rescue in disasters. In wireless networks, the communication capability of UAVs was exploited to use them as aerial base stations and relay nodes~\cite{7470933,khamidehi2019power, rahmati2019energy}. Recently, deployment of UAVs for data collection from a set of battery limited users has attracted a lot of interest (e.g.,~\cite{shakhatreh2018optimal,chen2019lifetime,khamidehi2019reinforcement}). 

In a wireless network, upon having nodes with power constraints, such as mobile users or low power wireless sensors, the efficient utilization of their batteries and prolonging their lifetime are important concerns. The
usage of UAVs to  increase the time duration of uplink  transmissions of wireless devices is a relatively new topic of study. In~\cite{shakhatreh2018optimal}, serving ground nodes via a  single  UAV  acting  as an aerial  base station (BS) is considered, where the problem  of  optimizing the  UAV  placement to  maximize  the  lifetime  of  wireless  devices is studied. In this work, the  lifetime is defined as the sum of time durations of uplink transmissions of the ground nodes. A similar problem is studied in~\cite{chen2019lifetime},  where the network comprises a set of ground  users and a set of UAVs, and optimize the UAVs' positions and the users' transmission powers are optimized to maximize the minimum lifetime of the ground users.
There exists another related body of literature, where the sensing protocol and UAV trajectories are designed to prolong the lifetime of the sensors (see~\cite{liu2019energy} and references therein). None of the above works considers the network lifetime maximization problem involving the battery consumption of the UAVs, where the UAVs act  as relays gathering information from ground sensors.

There exists another line of works  devoted to efficient \textit{leader} selection for the UAV swarms and the ground wireless devices. In~\cite{7106967}, using a UAV swarm for  the  scenario  of high-quality  forest mapping is investigated, where an adaptive and reliable network structure is desirable to maintain swarm connectivity and communicability. To this end, a  leader election  algorithm is applied to  a  set  of  micro  UAVs, where the leader gathers information from the swarm, leads the swarm to its destination, and responsible for  communicating  to  a BS. Also, effective usage of leader UAVs for the purpose of collision avoidance of UAV swarms is investigated in~\cite{8759940}, where each leader UAV is regarded as the controller and flies independently, and the others follow the leader UAV. In~\cite{8458023}, effective \textit{cluster head election} in UAV-assisted data collection in a  wireless sensor network is studied. To save  energies  and  extending  the  network  lifetime, the network is partitioned  into groups, each of which  that has a leader to communicate  communicates with the corresponding  UAV. Based on these works, we propose a more comprehensive model, in which the existence of a set of sensors and multiple terrestrial leaders and a set of UAVs with a UAV leader is considered simultaneously.

 The novelty and contributions of this work can be summarized as follows. First, we consider a new system model for data collection from a set of terrestrial sensors, where sensors are clustered to multiple groups each associated with a cluster head and a set of UAVs are deployed with a UAV leader for information gathering form the sensors. The leader UAV  (with a higher battery capacity and transmission power compared to the rest of UAVs) is in charge of data forwarding to  a base station (BS) through  backhaul network. Second, we formulate and solve the
 problem of obtaining the best positions of the collecting UAVs and the UAV leader so as to maximize the lifetime of the network while explicitly considering  the existence of multiple sources of interference in the environment.It is among the first work addressing the network lifetime maximization problem considering multiple jammers, and thus also contributes to the literature on UAV-assisted network design in the presence of jammers in the environment (e.g.,~\cite{One-leader,9013532,8761202,8737472,hosseinalipour2019interferenceJournal,rahmati2019dynamic2}).

\section{System Model}\label{sysmo}

We consider a scenario for UAV-assisted data collection, in which there are $N$ UAVs in the sky aiming at collecting information from $N$ clusters of terrestrial low power sensors. In each cluster, the sensors gather the required information and forward it to a primary  node, which we call \textit{cluster head} (CH). The aim of the CH is to collect information from the sensors in the same cluster and transmit it to the leader through UAVs. The main goal of the UAV network is to collect the data of the CHs and transmit it to a base station (BS). In practice, it is preferred that all the UAVs transmit their data to a high power UAV called \textit{leader} that relays the data to the main BS via a backhaul network.  The deployment of the leader is desired from the energy efficiency perspective since it results in prolonging the lifetime of the network.

In a detrimental scenario, the sensors and the UAVs coexist with multiple \textit{jammers}. The jammers' goal is  to generate strong interference to degrade the communication quality of the links from the CHs to the UAVs as well as the links from  UAVs to the leader. Nevertheless, the mobility feature of the UAVs allows them to reconfigure their locations and find their optimal 3D locations to avoid the interference caused by the jammer. In this work, we aim to optimize the location of the leader and also those of the data collecting UAVs to improve the lifetime of the network. In our work, we assume that the leader is a high altitude platform (HAP) which has sufficient power and can accommodate the transmission requirement from   the  data gathering UAVs  \cite{8660516}. Since we assume that the leader is a HAP, there should be a constraint on the altitude of the leader to prevent it from coming to the lower corridor in 3D space~(see Fig. 1 for more details). 
We propose a two step optimization method aiming to prolonging the lifetime of the network. In the first step,  we jointly optimize the locations of the data gathering UAVs and the leader to prolong the network operation lifetime. In the second step, given the location of the leader, the locations of the backhaul UAVs are optimized to maximize the lifetime of the backhaul network to deliver the data from the leader to the BS. {\color{black} Another interesting problem is to optimize the positions of the jammers, e.g.,~\cite{hosseinalipour2020optimal}, which we leave as a future work.}

 We denote the location of terrestrial jammer $j \in \mathcal{J}$ by $\mathbf{x}_j^{J} = (x_j^{J}, y_j^{J}, 0)$, the location of CH $m \in \mathcal{M}$ by $ \mathbf{x}_m^{C} =  (x_m^{C}, y_m^{C}, 0)$, and the location of  UAV $n \in \mathcal{N}$ by $\mathbf{x}_n =  (x_n, y_n, h_n)$.
In practice, the air to air (A2A) link, i.e., the inter-UAV link, is dominated by the line of sight (LoS) component, while the ground to air (G2A) link, i.e., the link between the CHs and the UAVs, can be considered as either LoS or non line of sight (NLoS). Here, we consider the worst case scenario in which the G2A link is assumed to be NLoS. The inverse of the  path-loss coefficients of the link between two nodes $p$ and $q$ is given by~\cite{khuwaja2018survey}:
\begin{equation}
\Gamma_{pq} \triangleq \begin{cases}
\displaystyle 1/(K_o ^{\alpha_1} \mu_{\textrm{LoS}}), & \textrm{if the}~ p~ \textrm{to}~ q~ \textrm{link is A2A}\\
\displaystyle 1/(K_o^{\alpha_2} \mu_{\textrm{NLoS}}), & \textrm{if the}~ p~ \textrm{to}~ q~ \textrm{link is A2G}.
\end{cases}
\end{equation}
where $\alpha_1$ and $\alpha_2$ are the path-loss exponents  for LoS and NLoS cases and $K_o=4 \pi f_c/c$, for which $f_c$ denotes the carrier frequency and $c$ denotes the speed of light. Also, $ \mu_{\textrm{LoS}}$ and $ \mu_{\textrm{NLoS}}$  denote the corresponding attenuation factors~\cite{mozaffari2017mobile}.

\begin{figure}[!t]
\vspace{3mm}
	\includegraphics[width=8.6cm,height=7.2cm]{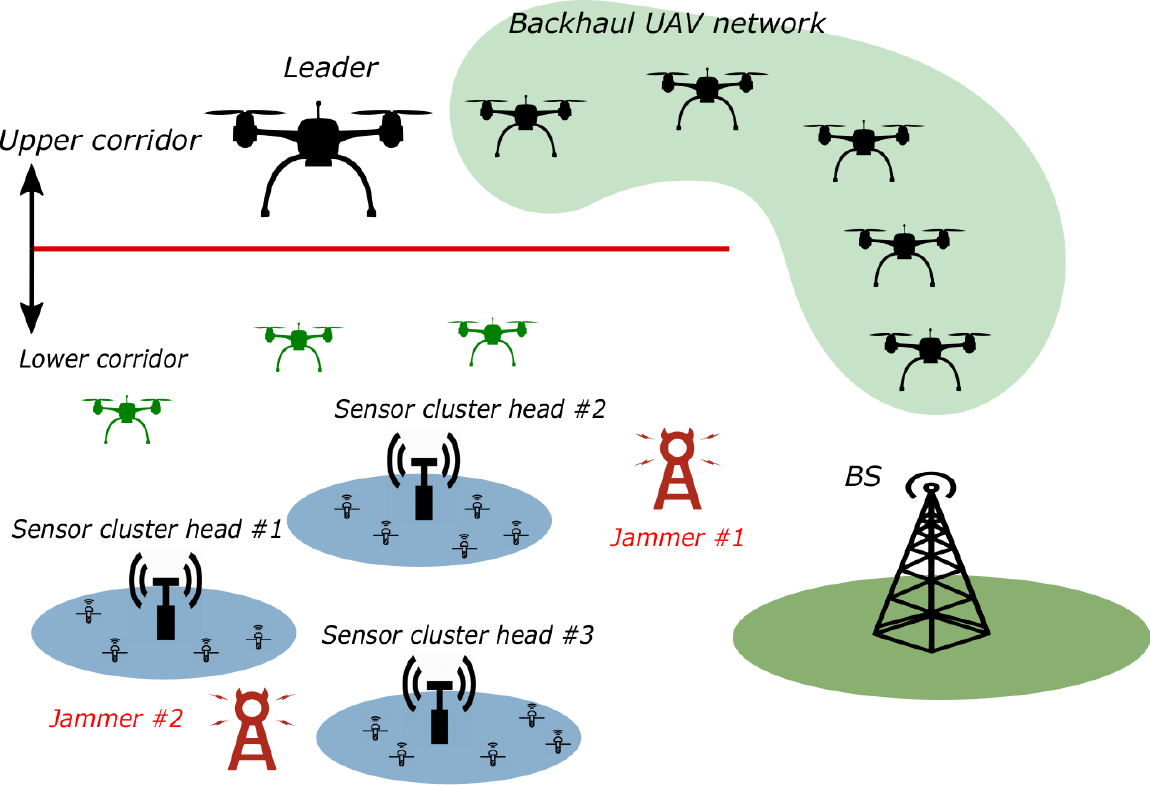}
	\centering
	\caption{The sample network with $N =3$ CHs/UAVs, one leader and $J=2$ jammers.}
\end{figure}
\section{Problem Formulation for UAV Lifetime Maximization}


\noindent We consider data transmission using the FDMA setting, where the UAVs utilize non-overlapping bandwidths to transmit data to the leader. For each UAV or CH $p$, $1 \le p \le N$, let $B_p = B/N$ denote the designated bandwidth. Consequently, the data rate from node $p$ to node $q$ is given by:
\begin{equation}\label{sir}
R_{pq}=B_p \log \left(1+ \frac{ P_{pq} \Gamma_{pq} d_{pq}^{-\alpha}}{ \sum_{j=1}^{J} P_j\Gamma_{jq} d_{jq}^{-\alpha}+\sigma^2} \right),
\end{equation}
where $P_{pq}$ is the transmit power required at  node $p$ to satisfy the transmission rate $R_{pq}$ at node $q$, $\sigma^2$ is the variance of the Gaussian noise, and  $d_{pq}$ denotes the Euclidean distance between the two indexed nodes and $P_j,~\forall j \in \mathcal{J}$ is the jammer transmission power.
Let $E_p$ denote the battery energy of node $p \in \mathcal{N} \cup \mathcal{M}$. We assume that $E_L$ denotes the battery energy of the leader,  where the battery energy of the leader is much larger than that of  the regular UAVs/CHs, i.e., $E_L \gg E_p, \forall p$. To maintain a certain quality of communication, each UAV/CH should be able to satisfy the minimum data rate requirement $R$ for the two-hop data transmission between the CHs and the leader through the UAVs. The leader, on the other hand, should be able to satisfy the transmission data rate of at least $N\times R$ to the backhaul network.  The lifetime of the link between two nodes $p$ and $q$ while delivering the minimum requirement rate $R_{pq}$ is given by:
\begin{equation}\label{lt}
    \tau_{pq}= \frac{E_p}{P_{pq}+P^c_{p}},
\end{equation}
where $P^c_{p}$ is a constant power consumption for each UAV/CH which is related to circuit power \cite{zhan2017energy}. Considering~\eqref{sir}, the transmission power $P_{pq}$ for node $p$ to satisfy the minimum rate $R$ at node $q$ is given by:
\begin{equation}
    P_{pq} = (2^{\frac{R}{B_p}}-1)\left({ \sum_{j=1}^{J} P_j\Gamma_{qj} d_{qj}^{-\alpha}+\sigma^2}  \right)\Gamma_{pq}^{-1}d_{pq}^{\alpha}.
\end{equation}
Since the leader is a high power HAP UAV, we assume that the bottleneck of the lifetime of the UAV network is determined by the data gathering UAVs and CHs. We define the lifetime of the network as the time duration that all of the links between the nodes can deliver a minimum data rate of $R$.
We formulate the optimization of the lifetime of the network as follows:
\begin{IEEEeqnarray}{rl}\label{opt1}
\max_{\substack{\mathbf{x}_p, \forall p \in \mathcal{N}, \mathbf{x}_L}} \qquad \hspace{-4mm} & \min_{\substack{p \in \mathcal{N} \cup \mathcal{M},~ q \in  \mathcal{N} \cup \{L\}}}  \qquad \tau_{pq} \label{nogiven} \\
\text{  s.t.}& 
\qquad  h_{\textrm{min}}  \le  h_L  , \IEEEyessubnumber \label{con1} \\
& ~~~~~~ d_{\textrm{min}}  \le d_{pq}  , \forall p, q \in \mathcal{N}, \IEEEyessubnumber \label{con222} 
\end{IEEEeqnarray}
where $\mathbf{x}_L$ is the location of the leader. In the above optimization, the objective function aims to maximize the minimum lifetime of the links between the nodes, which is equivalent to maximizing the lifetime of the network. Constraint \eqref{con1} ensures that the leader moves in the upper corridor and constraint \eqref{con222} guarantees that the UAVs observe a minimum safety distance $d_{\textrm{min}}$ with each other to guarantee the safety and proper flight requirements. The optimization is over both the locations of the regular UAVs and the leader. As can be seen, the location of the leader can have a high impact on the lifetime of the network. It can be verified that the objective function of optimization problem \eqref{opt1} is non-convex. 
The optimization in \eqref{opt1} is with respect to the locations of the data gathering UAVs and the leader, which is assumed as the first stage in our scheme. The second stage, which is optimization of the locations of  the backhaul UAVs  can be done similarly in a multi-hop setting, where the  leader  is fixed and acts as the source and the BS is the destination.
In such a case, the leader acts as source and the BS serves as the destination, while the backhaul UAVs serve as the relays.
\section{UAV Positioning using Spectral Graph Theory}\label{sec4}
\noindent The optimization problem \eqref{opt1} is a highly non-convex max-min problem. 
To address this problem, we deploy the \textit{Cheeger constant} of the given network, which is a well-known metric that measures the bottleneck of the network. 
\noindent We define a directed flow graph 
$G=(\mathcal{N},\mathcal{E})$, where $\mathcal{E}$ denotes the set of available edges in the network that correspond to the communications between the nodes. We assume that there exists an edge between node $p$ and  node $q$, where $a_{pq}$ denotes the weight of the edge connecting node $p$ to node $q$, which is defined as the lifetime of their corresponding link:
\begin{equation}
   a_{pq}= \frac{E_p}{(2^{\frac{R}{B_p}}-1)\left({ \sum_{j=1}^{J} P_j\Gamma_{qj} d_{qj}^{-\alpha}+\sigma^2}  \right)\Gamma_{pq}^{-1}d_{pq}^{\alpha}+P^c_{p}}.
\end{equation}
 The generalized adjacency matrix of the network is defined as $\mathbf{A}=[a_{pq}]$. 
Let us define the matrix $\mathbf{D}=\textrm{diag}\{\beta_1, ..., \beta_N\}$ as the generalized degree matrix of the UAV network with $\beta_p= \sum_{q, q \ne p}a_{pq}$. The Laplacian matrix of the network graph is defined as $\mathcal{L}=\mathbf{D}-\mathbf{A}$.
The weighted Cheeger constant is defined as:
\begin{equation}
h_{\mathbf{W}}(\mathcal{L}_\mathbf{W})=\underset{S}{\text{min}} \frac {\sum_{i \in S, j \in \bar S}a_{i,j}}{\text{min}\{ |S|_{\mathbf{W}},|\bar S|_{\mathbf{W}}\}},
\end{equation}
where the $|S|_{W}=\sum_{i \in S} w_{i}$ is the weighted cardinality  with $w_{i} \geq 0$ the weight assigned to node $i$,   $S$ is a subset of nodes and $\bar{S}$ denotes its complement. Also, $\mathcal{L}_\mathbf{W}$ denotes the weighted Laplacian defined as:
\begin{equation}
\mathcal{L}_\mathbf{W}=\mathbf{W}^{-1/2}\mathcal{L} \mathbf{W}^{-1/2}, \textrm{ with}~ \mathbf{W}=\textrm{diag}\{w_1, ...,w_n\}.
\end{equation}
Note that increasing the Cheeger constant of the network results in decreasing the bottleneck of the network, which is the aim of the optimization problem~\eqref{opt1}.
However, in general, instead of obtaining the Cheeger constant, the \textit{algebraic connectivity} $\lambda_2$ is computed as it is closely related to the Cheeger constant and is also simpler to compute.
It is shown in \cite{6807812} that the following weighted Cheeger's inequality holds
\begin{equation}\label{ine}
\lambda_2(\mathcal{L}_\mathbf{W})/2 \le h_\mathbf{W}(\mathcal{L}_\mathbf{W}) \le \sqrt{2 \delta_{\textrm{max}} \lambda_2(\mathcal{L}_\mathbf{W})/w_{\textrm{min}}} \,,
\end{equation}
where $\delta_{\textrm{max}}$ is the maximum node degree,
and $w_{\textrm{min}}= \min_i w_i$. 
In practice, since computing the Cheeger constant is difficult, we try to maximize  $\lambda_2({\mathcal{L}_\mathbf{W}})$ as an alternative according to \ref{ine}, which is given by:
\begin{equation}
\lambda_2(\mathcal{L}_\mathbf{W})= \underset{\mathbf{v}\ne \mathbf{0}, \mathbf{v} \perp \mathbf{W}^{1/2}\mathbf{1}}{\text{inf}} \frac{\langle \mathcal{L}_\mathbf{W} \mathbf{v}, \mathbf{v}\rangle}{\langle \mathbf{v}, \mathbf{v}\rangle}. 
\end{equation}

 
  \begin{figure*}[ht!]
  \subfloat[\label{fig1a}]{%
      \includegraphics[width=0.33\textwidth]{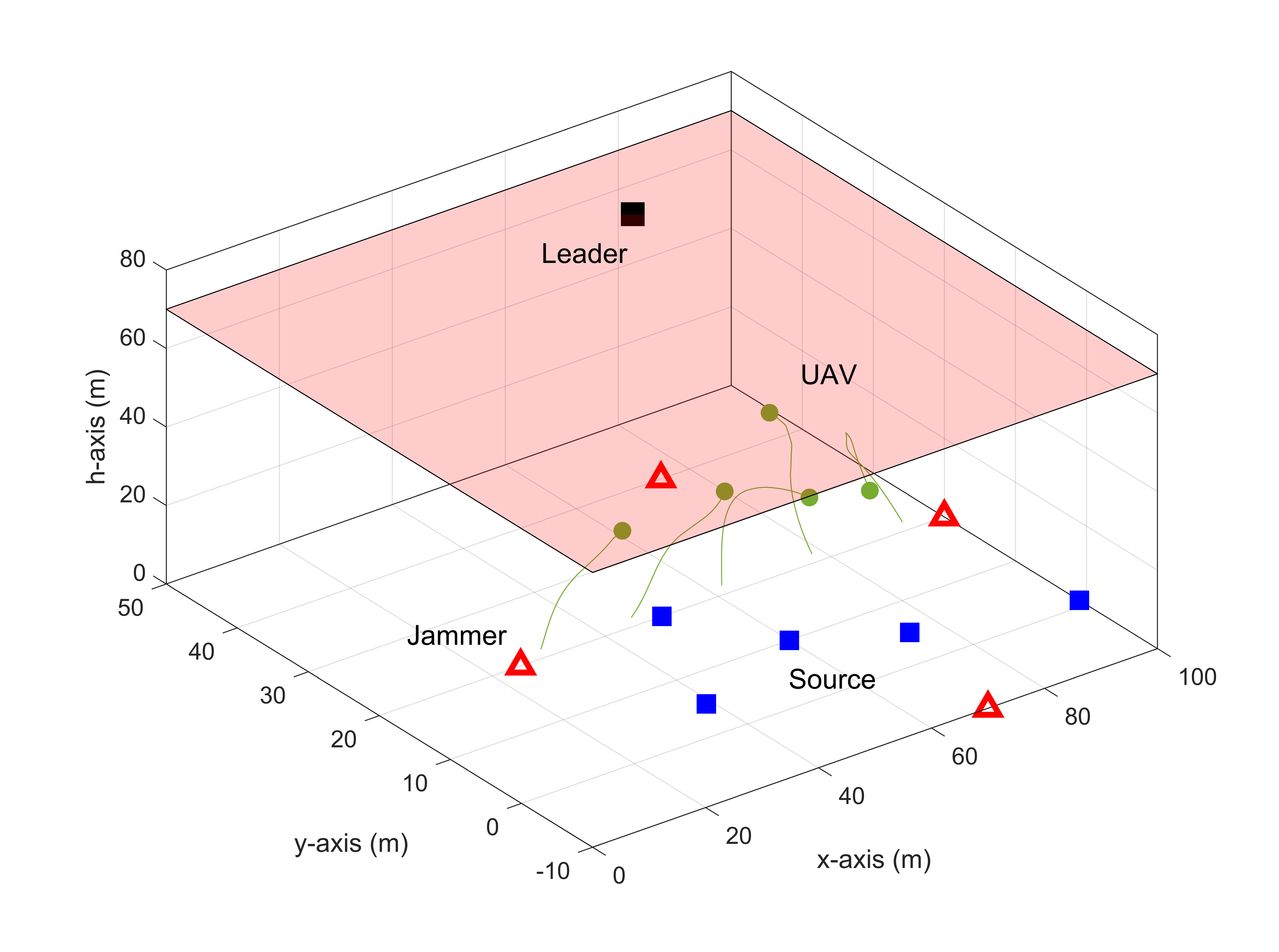}}
\hspace{\fill}
  \subfloat[\label{fig1b} ]{%
      \includegraphics[width=0.33\textwidth]{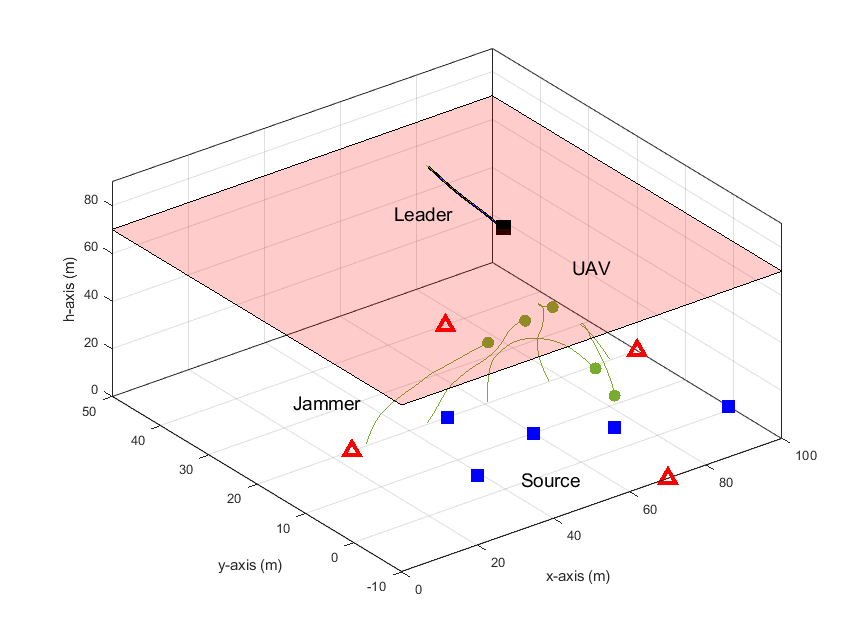}}
\hspace{\fill}
  \subfloat[\label{fig1c}]{%
      \includegraphics[width=0.33\textwidth]{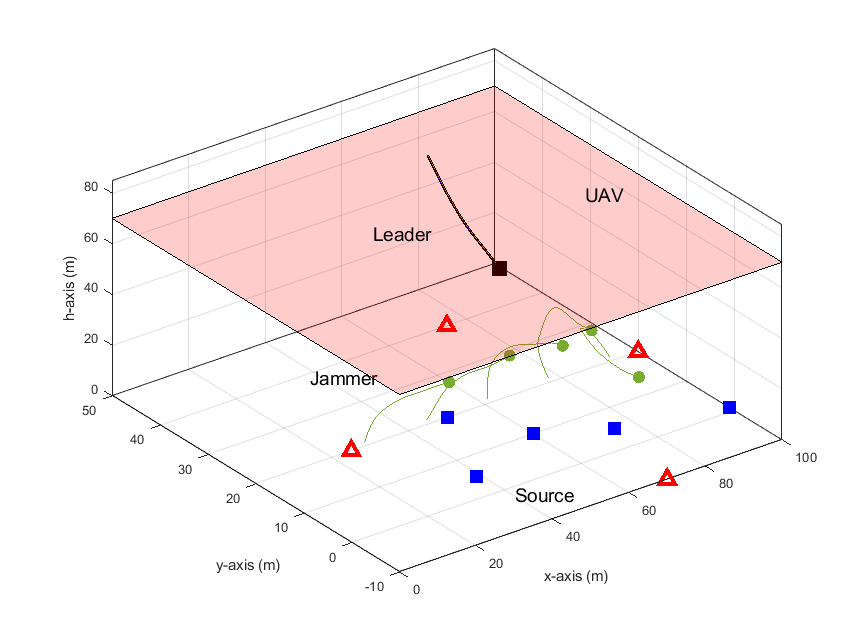}}\\
\caption{\label{costrr} The trajectories of the leader and the UAVs in the presence of 4 jammers with 5 UAVs, and 5 sensor cluster heads. (a)  The leader location is fixed on the upper corridor. (b) The leader movement is limited to the upper corridor. (c) The leader can freely move in the 3D space (Black square, green circles, blue squares, and red triangles  represent the leader, the UAVs, the sensor cluster heads, and the jammers, respectively.). }

\end{figure*}

 In order to maximize the weighted  algebraic connectivity $\lambda_2({\mathcal{L}_\mathbf{W}})$, we move each UAV along the spatial gradient of $\lambda_2({\mathcal{L}_\mathbf{W}})$. Given the instantaneous position of UAV $i$, its spatial gradient along $x$-axis is given by:
\begin{align}
\frac{\partial \lambda_2({\mathcal{L}_\mathbf{W}}) }{\partial x_i} & =    {\mathbf{x}^f}^T \frac{\partial ({{\mathcal{L}_\mathbf{W}}}) }{\partial x_i} {\mathbf{x}^f}   \nonumber  \\
 &= \sum_{\{p,q:p \sim q \}}  \left[\frac{x_p^f}{\sqrt{w_p}}-\frac{x_q^f}{\sqrt{w_q}}\right]^2\frac{\partial a_{pq}}{\partial x_i}, & 
\label{eq:moving}
\end{align}
where $\mathbf{x}^f$ is the Fiedler vector (i.e.,  the eigenvector corresponding to the second smallest eigenvalue), ${x}^f_p$  is the $p^\textrm{th}$  component of $\mathbf{x}^f$, and $p \sim q$ means that
nodes $p$ and $q$ are connected. To compute \eqref{eq:moving}, we need to obtain $\frac{\partial a_{p,q}}{\partial x_p}$, and it should be noted that if $p=q$, $\frac{\partial a_{p,q}}{\partial x_i}=0$.  Taking the derivative of the edge weight with respect to   $x_p$, which is a transmitter node, one can get:
\begin{align}
\frac{\partial a_{pq}}{\partial x_p}\hspace{-1mm}&= \hspace{-1mm}   \frac{E_p(2^{\frac{R}{B_p}}-1)\left({ \sum_{j=1}^{J} P_j\Gamma_{qj} d_{qj}^{-\alpha}+\sigma^2}  \right)}{\left(\hspace{-1mm} (2^{\frac{R}{B_p}}-1)\hspace{-1mm}\left({ \sum_{j=1}^{J} P_j\Gamma_{qj} d_{qj}^{-\alpha}+\sigma^2}  \right)\hspace{-1mm}\Gamma_{pq}^{-1}d_{pq}^{\alpha}+P^c_{p}\right)^{\hspace{-1mm}2}} \nonumber  \\
  \begin{split}
\times   \left[\alpha \Gamma_{pq}^{-1}d_{pq}^{\alpha-2}(x_q-x_p) \right].
    \end{split} & 
\label{eq:1}
\end{align}

Similarly, the derivative with respect to the position of the receiver  node $q$  is given by:
\begin{align}
\frac{\partial a_{pq}}{\partial x_q}\hspace{-1mm}&= \hspace{-1mm}   \frac{E_p(2^{\frac{R}{B_p}}-1)}{\left(\hspace{-1mm} (2^{\frac{R}{B_p}}-1)\hspace{-1mm}\left({ \sum_{j=1}^{J} P_j\Gamma_{qj} d_{qj}^{-\alpha}+\sigma^2}  \right)\hspace{-1mm}\Gamma_{pq}^{-1}d_{pq}^{\alpha}+P^c_{p}\right)^{\hspace{-1mm}2}} \nonumber  \\
 \begin{split}
\times   \left[\alpha \Gamma_{pq}^{-1} d_{pq}^{\alpha}\left({ \sum_{j=1}^{J} P_j\Gamma_{qj} d_{qj}^{-\alpha-2} (x_q-x_j)}  \right) \right. \\
    \left. + \alpha\Gamma_{pq}^{-1}d_{pq}^{\alpha-2}(x_q-x_p)\left({ \sum_{j=1}^{J} P_j\Gamma_{qj} d_{qj}^{-\alpha}+\sigma^2}  \right)\right ].
    \end{split} & 
\label{eq:1}
\end{align}
For coordinates $y$ and $z$, the moving directions can be obtained similarly, which are omitted due to space limitations. 

\section{Simulation Results}
\noindent In this section, we present a set of numerical simulations  to evaluate the performance of the proposed method. The simulation parameters are given in Table 1.
In Fig. 2, the 3D trajectories and final optimal locations (indicated by the corresponding markers) of the UAVs and the leader are shown, where   the first stage optimization for the locations of the data collecting UAVs and the leader is considered. The  hyperplane of $z=70$ m separates the upper corridor and the lower corridor.  In Fig.~2(a), the location of the leader is fixed in the upper corridor and only the locations of the UAVs are optimized in order to improve the lifetime of the network. In Fig.~2(b), the joint optimization of the locations of the UAVs and the leader is considered, while the moving space of the leader is constrained to the upper corridor.  Finally, in Fig.~2(c), the locations of the UAVs and the leader are jointly optimized while the leader can move freely in the 3D space.  This case can improve the overall performance in terms of lifetime maximization, while it violates the feasibility constraint. We consider this case as a potential upper bound for the performance of the network. In all of the scenarios, the locations of the data gathering UAVs is obtained using the proposed method. In the two latter cases, the black lines in the figures represent the trajectory of the leader.


\begin{table}[h]
\caption {Simulation Parameters} \label{tab:simulation} 
\renewcommand{\arraystretch}{1.2}
\centering
\begin{tabular}{ lc }
\hline
Parameter & Value \\
\hline
\hline
Number of data gathering UAVs & 5\\
Number of jammers& 4\\
Number of CHs& 5\\
Path-loss exponents $(\alpha_1, \alpha_2)$ & $2.05, 2.32$  \\
Transmit power of jammers $(\textrm{P}_j, \forall j)$ & $30\,\text{dBm}$ \\
Bandwidth $({B})$ & $10\,\text{KHz}$  \\ 
Safety distance $(d_{\textrm{min}})$ & $5\,\text{m}$ \\ 
Carrier frequency $(f_{\rm c})$ & $2~ \text{GHz}$ \\ 
Energy of each UAV/CH $(E_{p})$ & $20000 $ Joule \\ 
Data rate $(R)$ & $4 $ Mbps  \\ 
\hline
\end{tabular}
\end {table}
In Fig.~3, the lifetime of the network for the aforementioned cases is compared with a baseline method in which the UAVs are placed on the middle plane between the ground and the upper corridor. The locations of the CHs and jammers are chosen randomly in the area of $xy$ plane [0,100]$\times$[-10,40]. The results shown are averaged over 1000 realizations with $95\%$ confidence intervals. For the baseline method, the altitude of all the UAVs is $35$m, while the $x$ and $y$ positions of them are generated randomly. The location of the leader is assumed at the middle of the smallest rectangle in $xy$ plane including all the nodes with altitude 70m.  It can be seen that the proposed method can improve the lifetime of the network considerably. 
\begin{figure}[!t]
	\centering
	\includegraphics[width=0.45\textwidth]{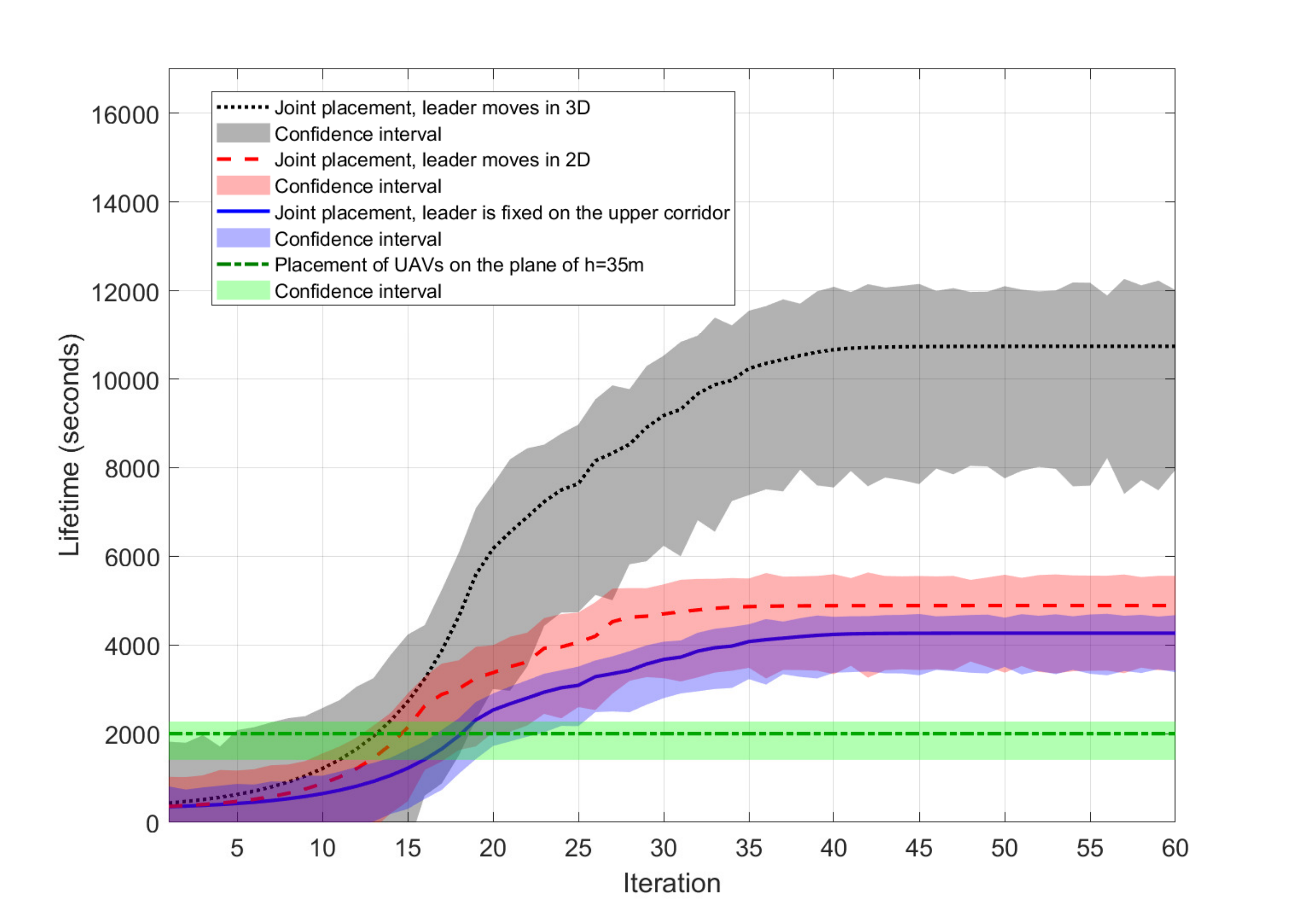}
	\caption{ Lifetime comparison of different scenarios.}
	\label{f2}
\end{figure}


The optimization of the locations of the UAVs in the backhaul network is also important to efficiently deliver the gathered information from the leader UAV to the BS. Given the location of the leader, one can obtain the  locations of the backhaul UAVs  using a similar approach. This scenario is depicted in  Fig.~4, where we optimized the positions of the leader, the data gathering UAVs, and the backhaul UAVs.

\section{Conclusion and Future Works}
\noindent 
In this paper, we investigated the problem of lifetime maximization of a UAV-assisted data gathering network, where UAV nodes collect data from a set of terrestrial low power wireless sensor nodes in the presence of sources of interference.  In the proposed scenario, the UAVs transmit all the collected data to a  UAV  \textit{leader}, which is in charge of forwarding the data to the main base station  via a backhaul UAV network. We formulated the lifetime maximization of the UAV network as an optimization problem, which turns out to be a highly non-convex max-min problem. To tackle the problem, we deployed Cheeger constant and algebraic connectivity from the spectral graph theory. Through numerical simulations, we studied the performance of our proposed method. There are many aspects of this problem that deserve further investigation including
sensor clustering, cluster head selection, and UAV grouping.

\bibliographystyle{IEEEtran}
\bibliography{IEEEabrv,references}
\begin{figure}[!t]
	\centering
	\includegraphics[width=0.4\textwidth]{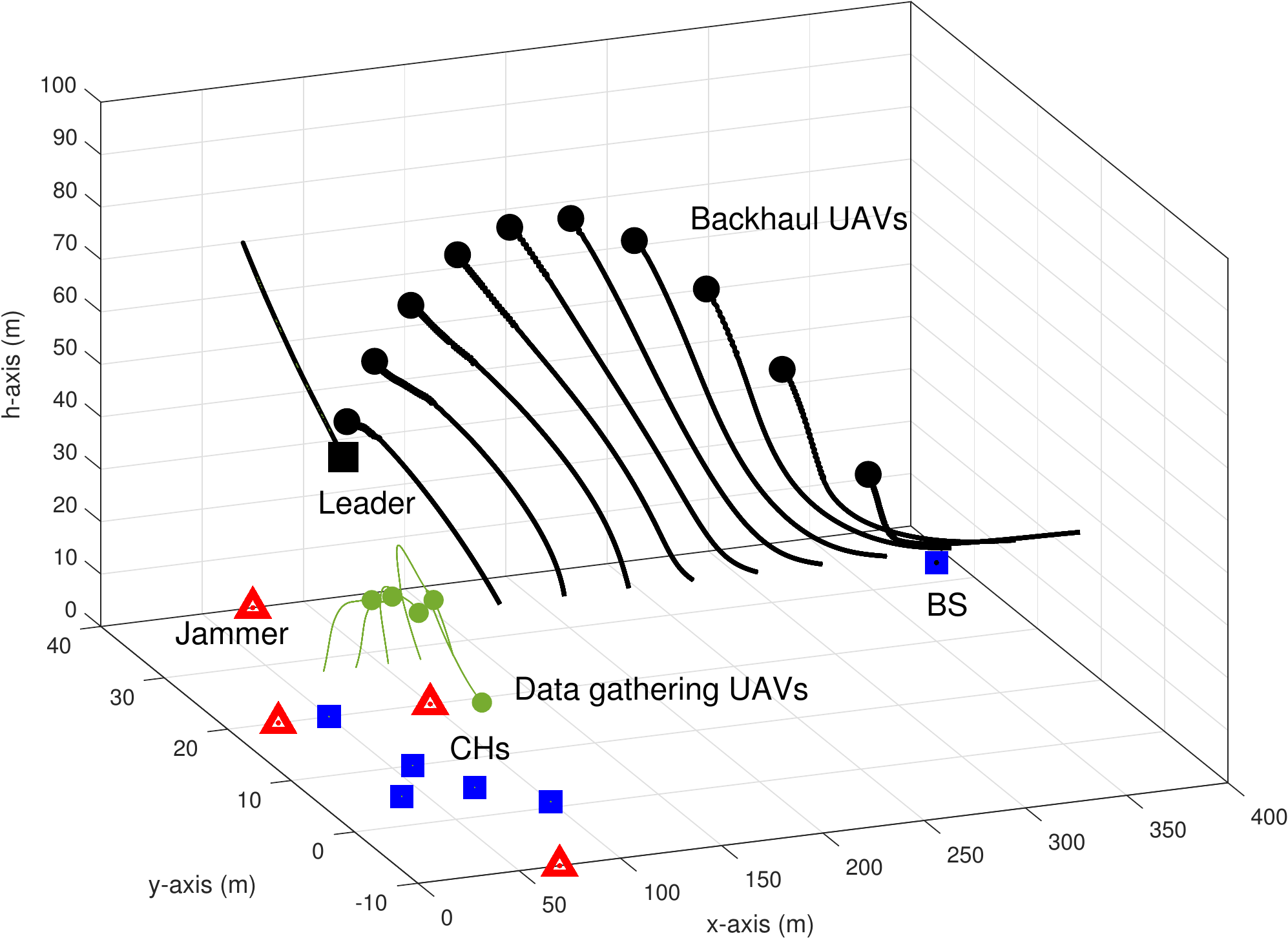}
	\caption{ Two step placement of the data collecting UAVs and backhaul network UAV. The black circles illustrate the optimized locations of the UAVs in the backhaul network.}
	\label{f2}
\end{figure}
\end{document}